\documentclass[aps,prf,superscriptaddress,preprint, amsmath,amssymb]{revtex4-2}

\usepackage{amsmath}
\usepackage{graphicx}
\usepackage{amssymb}
\usepackage{dcolumn}
\usepackage{bm}
\usepackage{color}

\makeatletter

\begin{document}

\title{Active chiral rotors: hydrodynamics and chemotaxis}

\author{R. Maity}
\affiliation{Department of Physics, Indian Institute of Science Education and Research, Bhopal 462066, India}
\affiliation{Institut f\"{u}r Theoretische Physik, Technischen Universit\"{a}t Wien, Wien 1040, Austria}

\author{Snigdha Thakur} \thanks{\tt Corresponding 
author:sthakur@iiserb.ac.in}
\affiliation{Department of Physics, Indian Institute of Science Education and Research, Bhopal 462066, India}

\author{P. S. Burada} \thanks{\tt Corresponding 
author:psburada@phy.iitkgp.ac.in}
\affiliation{Department of Physics, Indian Institute of Technology Kharagpur, Kharagpur 721302, India}

\date{\today}

\begin{abstract}

An active chiral rotor is a spherical object that can generate chiral flows in a fluid by rotating about an axis. For example, if the flow around the upper hemisphere of the chiral rotor is in a clockwise direction, then the flow in the lower hemisphere is in the anti-clockwise direction, and vice versa. 
In this paper, we aim to study the combined behaviour of hydrodynamically interacting chiral rotors in the presence of an external chemical gradient. While a single isolated rotor is stationary in a fluid, a pair of rotors can move in linear or circular paths as they hydrodynamically interact with each other. It is observed that the emergent linear or circular trajectories depend on the type of rotors and the orientation of their rotation axes. The dynamics of the rotors are altered in a more complex environment, such as in an external chemical field. Interestingly, we observe two types of motion: chemotaxis and anti-chemotaxis. While in the anti-chemotaxis case, both rotors are driving away from the target, in the chemotaxis case, one of the rotors successfully reaches the chemical target. This study helps to understand the collective behavior of self-propelled microorganisms and artificial swimmers.

\end{abstract} 

\maketitle

\section{Introduction}

Self-propulsion of microorganisms at a small length scale is ubiquitous in biological systems and has drawn considerable interest in the scientific community due to their swimming mechanisms and collective behaviors~\cite{ramaswamy, magistris,Eric,gompper}. 
Biological entities like bacteria, spermatozoa, and ciliated microorganisms are a few examples that exhibit very interesting individual as well as collective phenomena that can be associated with the surrounding fluid medium \cite{creppy,brumley,zhou}. 
Along with the translational motion, nature also provides us with ample examples of swimming 
microorganisms that display rotational movements e.g., \textit{Volvox}~\cite{drescher}, uniflagellar algae \textit{C.reinhardtii} ~\cite{guasto}, and \textit{T. majus} ~\cite{petroff}. Goldstein and co-workers reported stable bound states of spinning \textit{Volvox} ~\cite{drescher}. 
Similarly, the formation of rotating bacterial crystals by \textit{T. majus} was shown ~\cite{petroff} before.

Motivated by these natural swimmers, various synthetic self-propelled particles that exhibit directed motion have been designed to show intriguing collective features by hydrodynamic interaction among them ~\cite{theurkauff, yu, qiao}. While the literature on self-propelled systems exhibiting translational motion is rich, the studies on self-spinning particles (rotors) and their collective behaviour are limited ~\cite{kokot, snezhko, bricard, bricard_nat, grier, friese, wang, yamamoto_crystal}. 
Recent experimental realizations have successfully demonstrated such rotors by different mechanisms ~\cite{kokot, snezhko,bricard, bricard_nat,grier, friese,wang}. Chemical, light, and magnetic activation are a few examples that lead to the formation of such rotors ~\cite{zong, dong}. 
Although the rotor itself is an interesting entity to study, the presence of other rotors in the vicinity makes its dynamics more fascinating. A study on a collection of magnetic rotors has shown dynamic self-assembly due to the attraction and flow-induced interactions~\cite{climent, mecke} between them. Recent experimental work on bacteria–polymer mixture has demonstrated the self-assembly in active rotor suspension~\cite{linek}. Similar to these, in recent years, some interesting works emerged on rotor collective dynamics at interfaces \cite{llopis,furthauer,uchida} and in suspensions \cite{yeo_rheo, jibuti}.

In the low Reynolds number region, 
in the active rotor systems, long-range hydrodynamic interaction (HI) plays an essential role in the synchronization and self-assembly~\cite{mecke,llopis,uchida,leoni,llopis_epl,nguyen,shen, chen_11}. It has been shown that while an isolated rotor does not translate, a pair of rotors leads to the motion of their centers due to HI~\cite{leoni,fily, lushi, das}. So far, studies on the influence of hydrodynamics flowfield on neighbouring rotors have been primarily limited to pair rotations and have been shown to be complex. Cooperative self-propulsion~\cite{fily}, a combination of fast and slow motions~\cite{leoni}, and Quincke rotation~\cite{das} for an otherwise pinned state have been observed for such pairs. 
It must be noted that most of the above works were on torque-driven microrotors, whereas in nature, there are examples of chiral rotation. 
The flow field associated with the chiral rotors comprises rotlet dipoles where the flow field in the upper hemisphere is in the clockwise direction, and the lower hemisphere is in the counter-clockwise direction or vice versa. 
For example, such a chiral rotation has recently been observed in murine pancreas-derived organoids ~\cite{Frank_Pancreas}.
Chiral rotors can be considered as a potential candidate to mix and transport fluids \cite{campbell,ballard}. While the dynamics of an isolated chiral rotor are predictable, the pair dynamics can be quite complex and are poorly understood. Therefore, it is essential to understand the dynamics of rotors on a more fundamental level to interpret their various behaviors. 

Furthermore, biological or synthetic microswimmers \cite{kirchman, saintillan,brenner, hong_rotor,kumar,maityepje,maityjfm,maitypof,thakur,thakur_2} can sense the external stimuli and respond accordingly. For example, under the influence of a chemical field, a microswimmer may swim either towards (positive chemotaxis) or away (negative chemotaxis) from the chemical source. 
Chemotaxis is vital in many natural phenomena, for example, pattern formation and fertilization ~\cite{friedrichpnas}. 
In the literature, spinning \cite{qin_acs, wang_acs} or orbiting \cite{fattah, gibbs_2019, gibbs_2011, gibbs_2010} micromotors powered by a local chemical gradient induced by chemical fuel is common, but the study observing the motion of rotors in an external chemical landscape is limited. Orbiting rotors are helpful in scanning the periphery of sample cellular organisms, which could be beneficial for biomedical use. On the other hand, translating rotors could be used for targeted drug delivery to the biological tissues \cite{cheng_2014, wu_2018}. The rotor is a better chemotactic candidate to reach a diseased cell as its rotating power enhances its penetrating capability in complex medium \cite{wu_2018, schamel_2014, walker_2015}.

In this article, we study the dynamics of a pair of chiral rotors using the chiral squirmer model \cite{burada, maity} considering only rotational flow fields. Although an isolated rotor cannot move, a pair of rotors may show a combined behavior due to hydrodynamic interaction. 
Further, we study the response of a pair of chiral rotors to an external chemical gradient. 
Here, because of the interplay between hydrodynamic interaction and chemotaxis, rotors may exhibit interesting behaviors. 
This paper is organized as follows. In Section ~\ref{sec:model}, we introduce the chiral rotor model and its corresponding Stokes equation and solutions. 
The hydrodynamic behavior of a pair of chiral rotors is discussed in Section ~\ref{sec:HI}. 
The response of hydrodynamically interacting rotors to the external chemical gradient is presented in Section ~\ref{sec:Chemotaxis}. 
The main conclusions are presented in Section~\ref{sec:Concluion}.

\section{Chiral Rotor Model}
\label{sec:model}

\begin{figure}[htb!]
\centering
\includegraphics[scale=0.5]{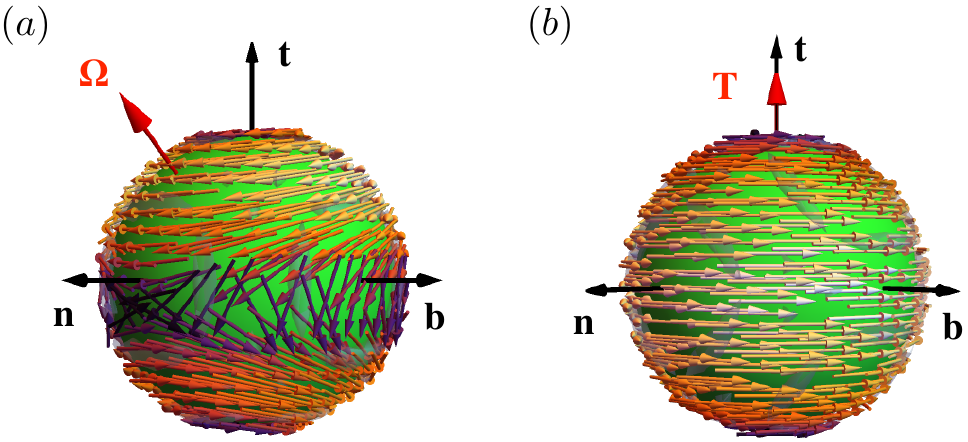}
\caption{$(a)$ Slip pattern of a chiral rotor in the body frame of reference $(\mathbf{n}, \mathbf{b}, \mathbf{t})$, i.e., Eq.~\ref{eq:slip}. 
Here, we consider $l = 1$ (with $m = 0,1$) and $l = 2$ (with $m = 0$) modes only.
The strength of $l=1$ mode chosen such that the rotation rate of the rotor $\mathbf{\Omega} = (\sqrt{3}/2,0,1/2)/a$ thus $v = a\,|\mathbf{\Omega}|= 1$. Similarly the strength of $l=2$ mode is set as $\gamma_{20}^r = 5\,v$. %The rotation rate is about $\bm \Omega$ axis. 
$(b)$ The flow pattern at the surface of a simple rotor (non-chiral) with an external torque $\mathbf{T}$ \cite{lushi} given by $v (a^3/8\pi\eta) \, \mathbf{T}\times {\bf \hat{r}}$.}
\label{fig:single rotor_flowfield sch}
\end{figure}

The flowfield around a low Reynolds number swimmer obeys Stokes equation,
\begin{align}
\nabla p = \eta \nabla^2 {\bf u} \, ,
\label{eq:Stokes}
\end{align}
where $\eta$ is the viscosity of the medium, ${\bf u}$ is the flowfield around the swimmer, and $p$ is the corresponding pressure field which plays the role of a Lagrange multiplier to impose the compressibility constraint $\bm{\nabla \cdot} {\bf u} = 0$. 
The model system, chiral rotor, is defined as a spherical body of radius $a$ with a prescribed chiral slip on the surface. It is similar to the chiral squirmer model \cite{burada}. 
The chiral squirmer can have both translational velocity and rotation rate, whereas the chiral rotor cannot swim on its own, but it can have only a rotation rate and generate chiral flows. 
The surface slip of the chiral rotor is defined in the body frame of reference $(\mathbf{n, b, t})$ (see Fig.~\ref{fig:single rotor_flowfield sch}) as \cite{burada}
\begin{align}
\label{eq:slip}
{\bf S}(\theta, \phi) & = \sum_{l=1}^{\infty} \sum_{m= -l}^l
\gamma_{lm}\, {\bf \hat{r}} \times {\boldsymbol \nabla}_s \left( P_l^m(\cos\theta) \,e^{i m \phi} \right), 
\end{align}
where ${\boldsymbol \nabla}_s$ is the gradient operator on the surface of the sphere defined as ${\boldsymbol \nabla}_s = {\bf e}_\theta \, {\partial}/{\partial}\theta +(1/\sin\theta)\, {\bf e}_\phi {\partial}/{\partial}\phi$, 
${\bf \hat{r}}$ is the unit vector in the radial direction measured from the center of the spherical body, 
$P_l^m(\cos\theta) \,e^{i m \phi}$
are non-normalized spherical harmonics, where $P_l^m(\cos\theta)$ denotes 
Legendre polynomials of order $m$ and degree $l$.
The complex coefficients $\gamma_{l m}$ are the mode amplitudes of the prescribed surface slip velocity. 
We introduce the real and imaginary parts of these amplitudes 
as $\gamma_{l m} = \gamma_{l m}^r + i\,m\, \gamma_{l m}^i$ 
with complex conjugates $\gamma_{l m}^\ast = (-1)^m \gamma_{l, -m}$, respectively.
The corresponding migration velocity and rotation rate of the chiral rotor can be obtained directly using the prescribed surface slip velocity (Eq.~\ref{eq:slip})\cite{Stone} as 
\begin{align}
\bm{V} & = \frac{-1}{4 \pi a^2}\int_{s} {\bf S}(\theta,\phi) \, ds = 0\, \\
\bm{\Omega} & = \frac{-3}{8 \pi a^3}\int_{s} \bm{\hat{e}}_r \times {\bf S}(\theta,\phi) \, ds = \frac{\gamma_{11}^r}{a}\mathbf{n} + \frac{\gamma_{11}^i}{a}\mathbf{b} + \frac{\gamma_{10}^r}{a}\mathbf{t} \, .
\end{align} 
%with $\omega = |\bm{\Omega}|$. 
We choose the rotation axis on $\mathbf{n}$-$\mathbf{t}$ plane for simplicity. 
Thus, $\gamma_{11}^i = 0$ throughout the current study. 
The angle between $\mathbf{t}$ axis and $\bm{\Omega}$ (see Fig.~\ref{fig:single rotor_flowfield sch}) 
is taken as $\chi$. Correspondingly, $\bm{\Omega} = (\sin \chi, 0, \cos \chi)/a$. 
We define $v = a|\bm{\Omega}| = a \,\omega$ as the characteristic velocity. 
In the following, we will alter the angle $\chi$ directly while studying the hydrodynamics of the chiral rotors.   

The surface slip of a chiral rotor is depicted in Fig.~\ref{fig:single rotor_flowfield sch}(a). 
The upper hemisphere moves in the clockwise direction, whereas the bottom one in the counter-clockwise direction. Since the symmetry in the slip pattern is broken, the rotor is called the chiral rotor. 
Accordingly, the chiral rotor generates asymmetric flows. 
However, in literature, a simple rotor (without chirality) is studied \cite{lushi, hong_rotor, lyu}, and is depicted in Fig.~\ref{fig:single rotor_flowfield sch}(b) for a comparison. In the latter case, the body spins about an axis and generates a simple rotational flow. 

The flow field of the chiral rotor can be calculated analytically by solving the Stokes equation (Eq.~\ref{eq:Stokes}) using the boundary conditions in the laboratory frame of reference are defined as 
\begin{equation}
  {\bf u}(\bm{r})=\begin{cases}
    {\bf S} + \left(\bm{V} + \bm{\Omega} \times \bm{\hat{e}}_r\right), & \text{at the surface of the rotor}.\\
    0 , & \text{at the far field}.
  \end{cases}
\end{equation}
The resultant velocity flow field reads
\begin{align}
\label{eq:VelocityField_rot}
{\bf u}_\mathrm{lf}({\bf r}) & =  
- \gamma_{2 0}^r\,\frac{a^3}{r^3} \, 
P_2^{\prime}\left( {\bf t} \cdot {\bf \hat{r}} \right)
{\bf t} \times {\bf \hat{r}} \,. 
\end{align}
where ${\bf u}_\mathrm{lf}({\bf r})$ is the velocity field in the laboratory frame of reference, 
$\gamma_{20}^r$ is the slip coefficient of the second mode corresponding to the rotational movement of the rotor, 
$r$ is the distance from the center of the swimmer where the flow field is determined, 
$P_2(x)$ denotes a second-order Legendre polynomial, and 
$P_2^{\prime} = dP_2/dx$ with $x = \mathbf{t}\cdot \mathbf{\hat{r}} = \cos \theta$.
Note that in Eq.~\ref{eq:VelocityField_rot}, the higher order terms $l > 2$ are being ignored as their contribution is negligible in the current study.
To have a minimal model, in Eq.~\ref{eq:VelocityField_rot}, $l=2$ modes with $m\neq 0$ have been ignored. 
The force- and torque-balance equations of the chiral rotor lead to  
\begin{align}
\dot{\mathbf{q}} &= 0 \, , \dot{\mathbf{n}} = {\bm \Omega} \times \mathbf{n} \, ,  \dot{\mathbf{b}} = {\bm \Omega} \times \mathbf{b} \, ,  \dot{\mathbf{t}} = {\bm \Omega} \times \mathbf{t},
\end{align}
where $\mathbf{q}$ is the position of the swimmer. 
Note that the single chiral rotor does not exhibit any translational motion, the time derivative $\dot{\mathbf{q}} = 0$.

\section{Hydrodynamic interaction of two chiral rotors}
\label{sec:HI}

\begin{figure}[htb!]
\centering
\includegraphics[scale=0.4]{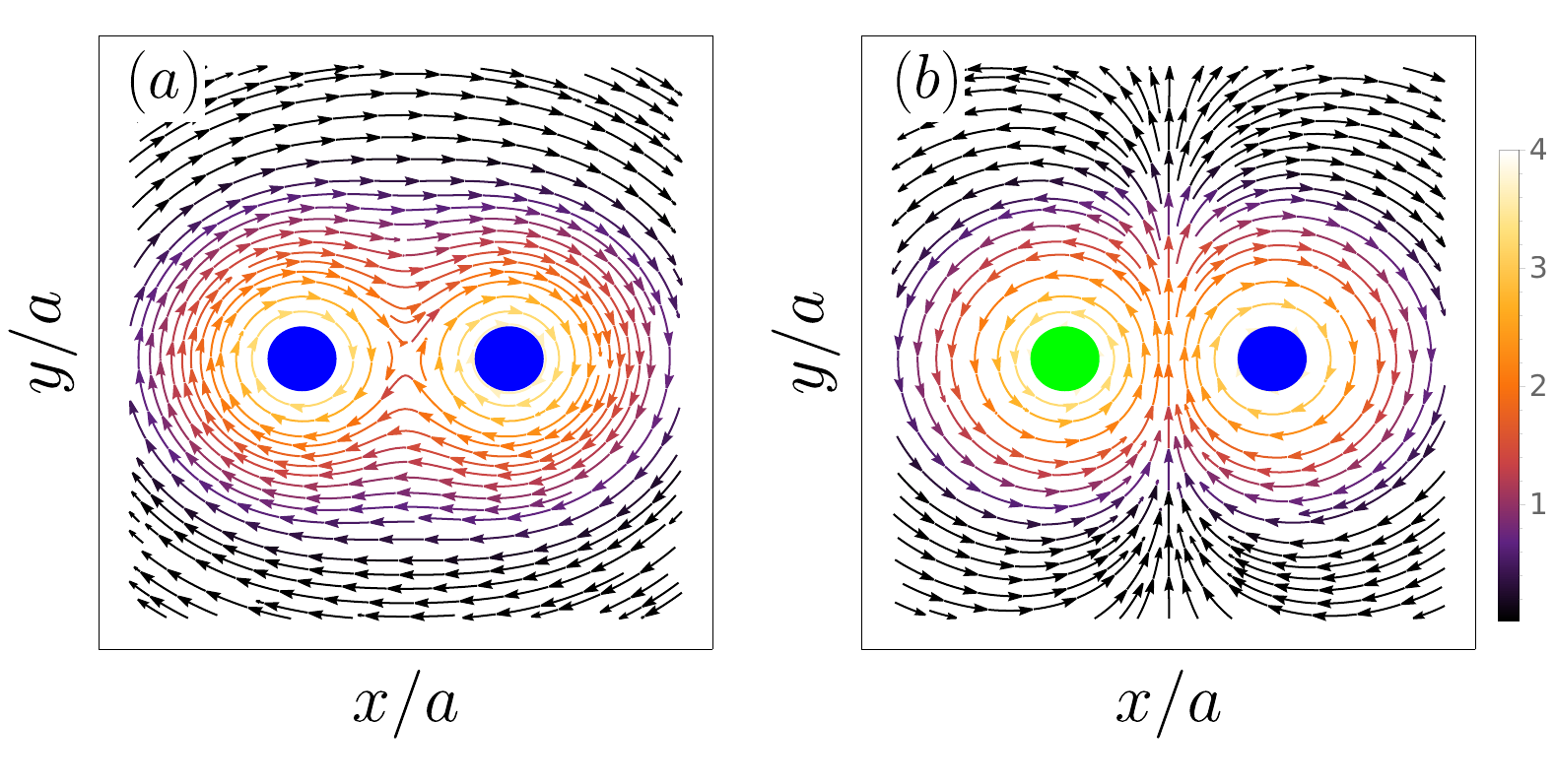}
\caption{The resultant flow field generated by (a) pair of rotors with the same chirality (blue spheres) situated at (9,3,0) and (3,3,0) with $\lambda/v = 5$, and,  (b) pair of rotors with opposite chirality (blue and green spheres) situated at (9,3,0) and (3,3,0) with $\lambda/v = 5$ and $-5$ respectively. 
The color bar implies the strength of the flow field.}
\label{fig:comb rotor_flowfield}
\end{figure}

When we place a second rotor near the first rotor, the flow field of one rotor influences the movement of the other and vice-versa. The resultant flowfields of a pair of rotors depend on 
the flow patterns of individual rotors, i.e., clockwise or anti-clockwise (see Fig.~\ref{fig:comb rotor_flowfield}). 
The equations of motion of a pair of rotors, in the presence of hydrodynamic interaction, read 
\begin{align}
\dot{\mathbf{q}}_i & = \sum^2\limits_{\substack{j=1 \,;\, i\neq j}}\mathbf{u}_j \,, \nonumber \\  
\dot{\mathbf{n}_i} & = {\bm \Omega}_i \times \mathbf{n}_i \, ,  \dot{\mathbf{b}_i} = {\bm \Omega}_i \times \mathbf{b}_i \, ,  \dot{\mathbf{t}_i} = {\bm \Omega}_i \times \mathbf{t}_i \, , 
\label{eq: rotors}
\end{align}
where $i=1$ for rotor one and $i=2$ for rotor two.
In general, the vorticity field $\nabla \times \mathbf{u}$ corresponding to the flow field of the rotor contributes to the above torque balance equation. However, in the present case, it is zero (see Eq.~\ref{eq:VelocityField_rot}).
Note that, for rotors, intrinsic translational velocities, $|\mathbf{V}_1| = 0 = |\mathbf{V}_2|$ and the rotation rates are ${\bm \Omega}_1 = (\sin \chi_1, 0, \cos \chi_1)/a$ and ${\bm \Omega}_2 = (\sin \chi_2, 0, \cos \chi_2)/a$. 
Furthermore, we set the slip coefficients of two rotors $\gamma_{20}^{r(1)} = \lambda= \gamma_{20}^{r(2)} = \lambda$.  
For the positive values of $\lambda$, the rotor generates the flow field that rotates in the clockwise (CW) direction. In contrast, for the negative values of $\lambda$, the flow field rotates in the anticlockwise (ACW) direction.

\begin{figure}[htb!]
\centering
\includegraphics[scale=1.4]{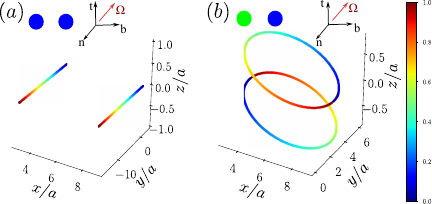}
\caption{(a) The two rotors with the same chirality (represented by blue spheres) move in a straight line parallel to the $y$-axis. Here, $\lambda/v = 5$ and $\chi = \pi/24$ for both the swimmers.
(b) Two rotors with the opposite chirality (represented by blue and green spheres) move in circular paths. Here, $\lambda$ and $\chi$ are the same as (a). The position of the rotor is the same as Fig.~\ref{fig:comb rotor_flowfield}. The normalized time is coded in the color bar.}
\label{fig:stagnant path}
\end{figure}

The resultant flow field by a pair of rotors about $z$-axis with the same and different chirality in the lab frame is plotted in Fig.~\ref{fig:comb rotor_flowfield}. Note that the $l = 1$ mode of Eq.~\ref{eq:slip} does not contribute to the flow field, but $l=2$ is the only responsible mode for the resultant flow field associated with the rotlet in the fluid. While the flow field corresponding to a pair of rotors with the same chirality adds up to give rise to a resultant clockwise flow field, the flowfields corresponding to a pair of rotors with different chirality add up to give rise to a resultant flow field where clockwise and anticlockwise vortices are merging. Hence, the observed trajectories of the rotors may be very distinct in these two cases. The pair of rotors with the same chirality (CW-CW or ACW-ACW) move in a straight line path either along the positive or negative $y$-axis depending on the strength of the resultant flow field (see Fig.~\ref{fig:stagnant path}(a)). Whereas, for dissimilar chirality of rotors (CW-ACW, ACW-CW), we observe circular trajectories (see Fig.~\ref{fig:stagnant path}(b)). Notably, we have chosen identical rotors, i.e., rotors with identical rotation rate, $\bm{\Omega}_1 = \bm{\Omega}_2 = (\sin \chi, 0, \cos \chi)/a$ with $\chi = \pi/24$ and identical flowfield strength, $\lambda_1 = \lambda_2$.

\begin{figure*}[htb!]
\centering
\includegraphics[scale=0.6]{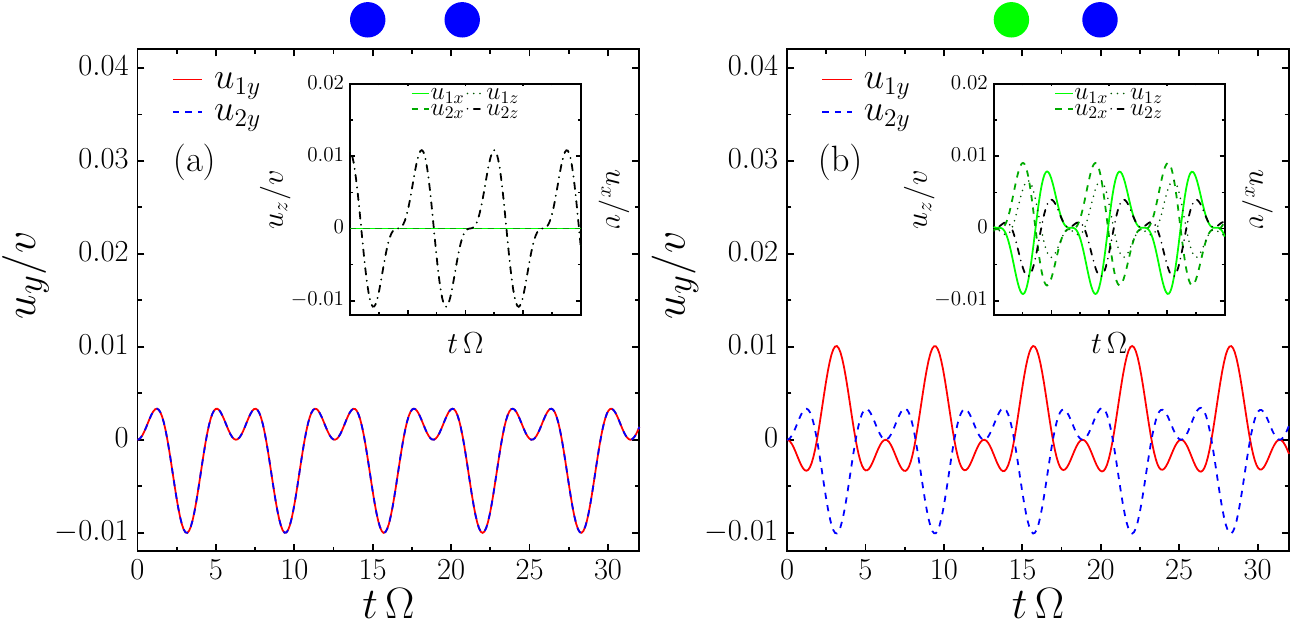}
\caption{The components of the flow field $(\mathbf{u})$ generated by one rotor at the position of the other rotor as a function of time, (a) for straight-line trajectories, and (b) for circular trajectories. All the corresponding parameters are the same as in Fig.~\ref{fig:stagnant path}.}
\label{fig:cw cw rotor_flowfield}
\end{figure*}

The quantitative analysis of the flow behavior sheds light on the nature of trajectories observed in the case of similar and dissimilar rotors. To see the hydrodynamic influence of one rotor on the other, in Fig.~\ref{fig:cw cw rotor_flowfield}, we plot the instantaneous flow field components $u_{1x,1y,1z}$ (for rotor one) and $u_{2x,2y,2z}$ (for rotor two) generated by a rotor at the position of the other. 
Fig.~\ref{fig:cw cw rotor_flowfield}(a) corresponds to the case where the rotors have the same chirality and move in straight-line trajectories. One can see the corresponding flow field components of the two rotors to be in phase for this case, indicating a synchronous movement between them. Note that the $u_{1x,2x}$ components are insignificant compared to the other two components, hence no movement in $x$-direction. Even though the instantaneous $y$ and $z$ components have comparable strengths, the time average of the $u_{1z,2z}$ is nearly zero. Hence, the rotors do not move in $z$-direction either. The net non-zero time-average values of $u_{1y,2y}$ make the rotors translate along the positive or negative $y$ direction depending on the values of $\chi$. 
However, the corresponding flow fields are out of phase for the dissimilar chiral rotors where we observe circular trajectories. The components $u_{1x,1y,1z}$ are phase shifted to $\pi/2$ with respect to $u_{2x,2y,2z}$ (see Fig.~\ref{fig:cw cw rotor_flowfield}(b)). 
Further, unlike the straight-line trajectories, all the components of the flow field contribute significantly to the dynamics. Therefore, the corresponding flow field components of both the rotors being out of phase sets them in circular trajectories.

\begin{figure}[htb!]
\centering
\includegraphics[scale=2.75]{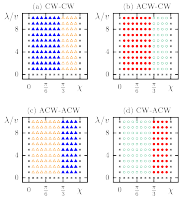}
\caption{The state diagram of the trajectories followed by the rotors for different combinations of their chirality, normalized strength of flow field $\lambda$), and $\chi$. $(a)$ The rotors have a clockwise (CW) flow field. $(b)$ The first rotor has an anti-clockwise (ACW) flow field, and the second rotor has a clockwise (CW) flow field. $(c)$ Both the rotors have anti-clockwise (ACW) flow fields.  $(d)$ The first rotor has a clockwise (CW) flow field, and the second rotor has an anti-clockwise flow field (ACW). Triangles denote the straight line trajectories where solid symbols are for motion in the $-ve \, y$ direction, while hollow symbols are for the $+ve \, y$ direction. Circles denote the circular trajectories of the rotors, with filled symbols for CW rotation and hollow symbols for ACW rotation. The symbol cross is for the stationary states of the rotors, i.e.,  no movement of the rotors.}
\label{fig:phase space two}
\end{figure}

The above discussion of the straight-line or circular trajectories was based on a particular choice of $\lambda$ and $\chi$. However, it will be interesting to see how the dynamics of the rotor change by changing these two parameters. To do so, we plot the state diagram of the rotors for various combinations of $\lambda$ and $\chi$ in Fig.~\ref{fig:phase space two}. 
As evident from this state diagram, rotors of the same chirality (same sign of $\lambda$) move in straight-line trajectories (triangle symbols) whereas rotors of opposite chirality (opposite sign of $\gamma_{2 0}^r$) move in the circular trajectories (circle symbols) for all $\lambda$ and $\chi$ (except $\chi = 0$ and $\pi/2$). For $\chi = 0$ and $\pi/2$, rotors do not influence each other, and as a result, there is no movement. 
This is because the magnitude of the generated flow field in these cases is not significant to translate the rotors. 
Note that the directionality of the motion is controlled by the parameter $\chi$. The state diagram shows a sharp change in the direction of motion near $\chi \approx \pi/3$ irrespective of the combinations of rotors, i.e., CW or ACW (see Fig.~\ref{fig:phase space two}). Interestingly, this direction flipping does not depend on the strength of the flow field $\lambda$.

\begin{figure}[htb!]
\centering
\includegraphics[scale=0.5]{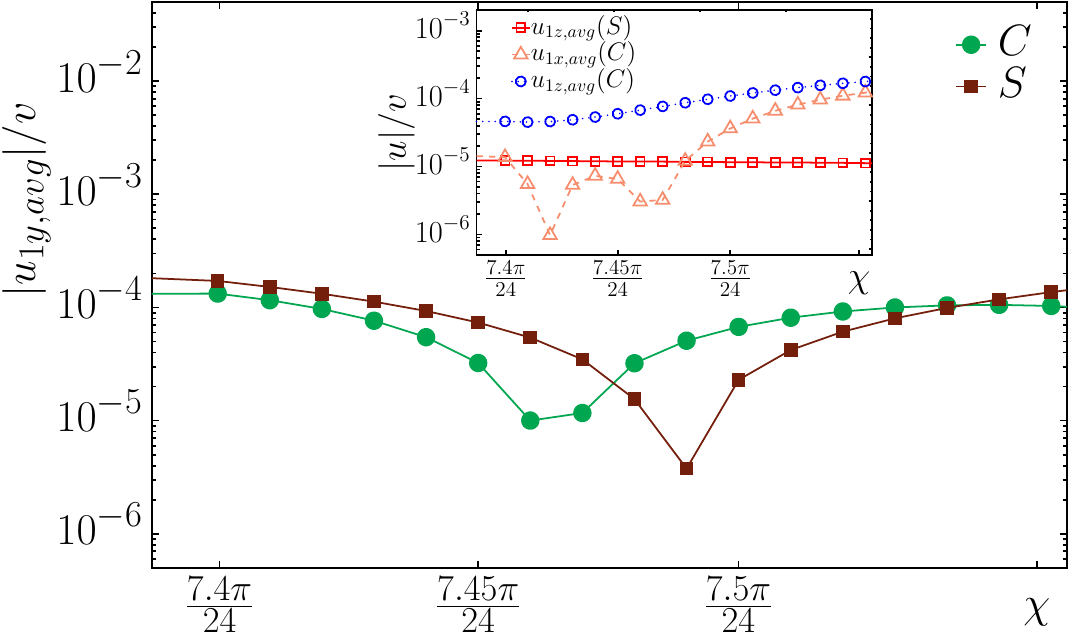}
\caption{The absolute value $\langle u_{1y} \rangle$  as a function of $\chi$ in a narrow range where the flipping of the trajectories is observed. $C$ stands for circular trajectory, and $S$ stands for the straight-line motion of the rotors.
The inset shows the variation of other components of the flow field.}
\label{fig:u1 chi r0}
\end{figure}

To comprehend this flipping of direction, we plot the absolute average value of the components of the flow field, i.e., $u_{1y}$ and $u_{1z}$ in Fig.~\ref{fig:u1 chi r0} for $\chi$ around $\pi/3$. This enables us to narrow down the search for flipping. It is clear that for both circular (C) and straight (S) trajectories, the $u_{1y}$ and $u_{1z}$ changes the sign near $\chi \sim 7.5 \pi/24$. It is this change of sign in the flow field that results in the motion reversal of the rotors.

\begin{figure*}[htb!]
\centering
\includegraphics[width=\linewidth]{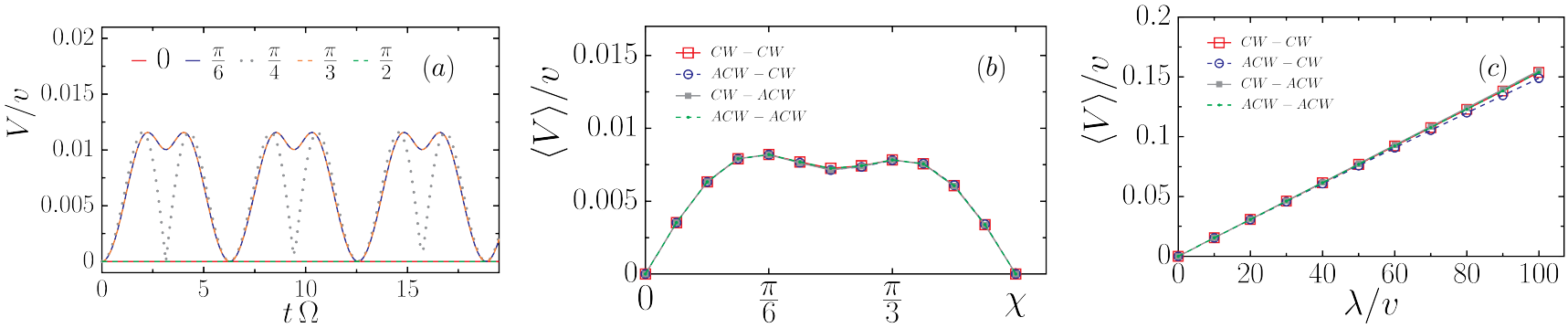}
\caption{$(a)$ Instantaneous speed $(V(t))$ of one rotor as a function of time $(t)$ for different $\chi$ values. The initial conditions and $\lambda$ are the same as Fig.~\ref{fig:stagnant path}. A similar curve is obtained for both combinations of chirality. $(b)$ Steady state root mean square velocity $(V)$ as a function of $\chi$ ranging from $0$ to $\pi/2 $ for combination of rotors presented in Fig.~\ref{fig:stagnant path}. All the corresponding parameters have the same values as in Fig.~\ref{fig:stagnant path}. Note that $\lambda = 5v$ for both the rotors. $(c)$ Steady state root mean square velocity $(V)$ as a function of $\lambda$ for combination of rotors presented in the Fig.~\ref{fig:stagnant path} with $\chi = \pi/3$. All the corresponding parameters have the same values as in Fig.~\ref{fig:stagnant path}.}
\label{fig:vel circle comb}
\end{figure*}

To get more insight into the rotor's translation, we calculate the instantaneous speed of the individual rotor ($V(t)$) in circular motion by finding the magnitude of the flow field imposed by the other rotor on it. That is the instantaneous speed of the first rotor $V_1(t) = \sqrt{u_{2x}^2 + u_{2y}^2 +u_{2z}^2}$, and similar for the other rotor. The instantaneous speed of one rotor is plotted in Fig.~\ref{fig:vel circle comb}(a), and due to the identical nature of the rotors, the speed of the other will also be the same. As evident from Fig.~\ref{fig:vel circle comb}(a), the rotors with $\chi = 0$ and $\pi/2$ (rotation about {\bf t} and {\bf n} axis, respectively) do not move as mentioned before, the magnitude of the generated flow field in these cases is not significant to contribute to the translational movement of the rotors. 
From the instantaneous speed, we calculate the average speed $\langle V \rangle$ of the rotor and plot it in Fig.~\ref{fig:vel circle comb}(b). The nature of $\langle V \rangle$ is symmetric about $\chi = \pi/4$. That is, the rotor has minimum speed at $\chi = \pi/4$. Interestingly, as shown in Fig.~\ref{fig:vel circle comb} (b), $\langle V \rangle$ in straight line trajectory is very similar to that of the circular trajectory. Note that the value of $\chi$ where the motion reversal happens is not the same as where $\langle V \rangle$ is minimum. This is because the symmetric part of the flow field ($x$ and $z$-components) does not contribute to the motion reversal (see Fig.~\ref{fig:cw cw rotor_flowfield}). Rather, only the asymmetric $y-$ component governs the motion reversal. However, all the components contribute to determining $\langle V \rangle$.

\begin{figure}[htb!]
\centering
\includegraphics[scale=1]{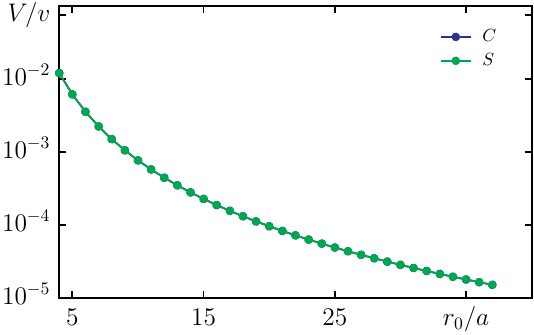}
\caption{The rms velocity field of the first rotor ($u_{1,rms}$) as a function of a varying initial separation distance $r_0/a$, here $\chi = \pi/3$. $C$ stands for circular paths, $S$ stands for straight line paths of rotors. The rest of the parameters are the same as in Fig.~\ref{fig:cw cw rotor_flowfield}.}
\label{fig:u1 r0}
\end{figure}

As the transnational motion of the rotors is entirely governed by the strength of the flow field of the neighbouring rotor, next, we see the effect of $\lambda$ (see Eq.~\ref{eq:VelocityField_rot}) in Fig.~\ref{fig:vel circle comb}(c). As expected, $\langle V \rangle$ varies linearly with $\lambda$ for all the rotor combinations.
Notably, the vorticity $(\nabla \times \mathbf{u}_{lf})$ goes as $\sim 1/r^4$ corresponding to the flow field (Eq.~\ref{eq:VelocityField_rot}), therefore the rotation rate $(\Omega)$ of the rotors is weakly influenced as the distance between them is large. Fig.~\ref{fig:u1 r0} shows the average speed $\langle V \rangle$ of one rotor as a function of the separation distance between the rotors. Indeed, at a distance $ \sim 40\,a$, the flow field strength $\sim 10^{-5}$ gives rise to insignificant displacement of the rotors. Henceforth, the rotors do not sense each other presence in the former case.

\section{Chemotaxis}
\label{sec:Chemotaxis}

\begin{figure}[htb!]
\centering
\includegraphics[scale=0.4]{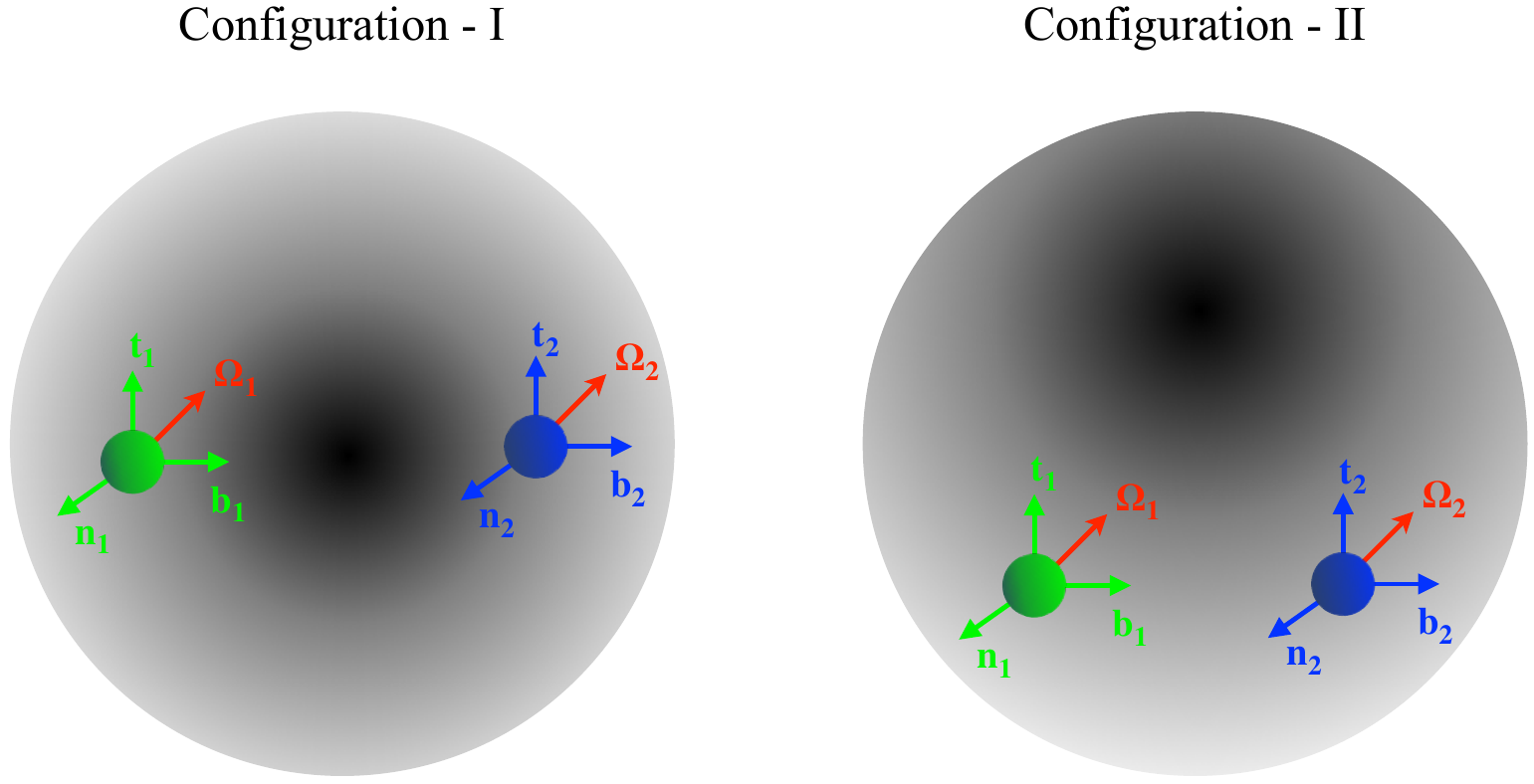}
\caption{Schematic diagram depicting a pair of rotors 
in a radial chemical field. Two configurations (I \& II) have been considered. I refer to co-linear with a source at the middle of the rotors, whereas II refers to a source off the centerline of the rotors.}
\label{fig:chem rotors}
\end{figure}

After understanding the combined behavior of two rotors, it is interesting to ask how they respond 
to external perturbations. In particular, understanding the interplay between hydrodynamic interaction and external field is important. 
In general, the external field alters the swimming characteristics of the rotors. 
For example, in some of the ciliated microorganisms, the surface of the body is covered with chemo-receptors \cite{alexandre}, and in the presence of a chemical gradient, chemoattractant molecules bind and unbind with the receptors sequentially, which are known as activation and deactivation respectively \cite{bradshaw, goy}. 
The former processes activate the internal chemical network, which modifies the slip coefficients of the rotor \cite{friedrichpnas}.
In this study, we have used the popular Barkai-Leibler model \cite{barkai,friedrichpnas,maityepje,maityjfm,maitypof} to model the adaptation and relaxation mechanism of the rotor in the presence of a chemical field. 
The following set of equations captures the swimmer's ability to adapt and respond to the weak chemical stimulus
\begin{align}
\sigma \dot{a}_b &= p_b(s_b + s) - a_b\, ,\label{rel}\\
\mu \dot{p}_b &= p_b (1-a_b)\, ,
\label{adap}
\end{align}
where $\sigma$ and $\mu$ are the relaxation time and adaptation time, respectively, $a_b(t)$ is the relaxation parameter, $p_b(t)$ is the adaptation or dynamic sensitivity, $s = c(\mathbf{q}(t))$ is the chemical stimulus, and $s_b(t)$ is due to the background activity of the receptor without chemical stimulus. %The Eqs.~\ref{rel} \& ~\ref{adap} are limited to a weak chemical gradient. 
$s$ has a dimension of concentration. In presence of a constant stimulus, $s(t) = S_c$, $a_b = 1$ and $p_b = 1/(s_b + S_c)$. Therefore, $p_b(t)$ inversely depends on the stimulus level $S_c$. As $a_b(t)$ is independent of $S_c$, the system is adaptive.
In the presence of an external chemical field, the slip coefficients are altered as,
\begin{align}
\gamma_{lm} &= \gamma_{lm}^0 + \gamma_{lm}^1(a_b(t) - 1)\,,
\label{eq:gamma_chem}
\end{align}
where $\gamma_{lm}^0$ are the unperturbed slip coefficients and $\gamma_{lm}^1$ are the perturbed slip coefficients. Here, we consider the case of rotors in the presence of an external radial chemical field defined as
\begin{align}
c(\mathbf{q}) &= c_r/r\, ,
\end{align}
where $c_r$ is the chemoattractant diffusivity, i.e., releasing rate of chemoattractant from the source $(x_0,y_0,z_0)$, and $r = \sqrt{(x-x_0)^2 + (y-y_0)^2 + (z-z_0)^2}$, is the distance between the rotor $(x,y,z)$ and the chemical source. 

Using Eq.~\ref{eq: rotors} we numerically simulate the swimming paths of a pair of rotors interacting hydrodynamically in a radial chemical field. Notably, the rotation rate of the rotors is a function of slip coefficients, which are modified due to the chemical gradient (Eq.~\ref{eq:gamma_chem}). The same holds for the flow field around the rotors. Thus, the interaction between the rotors is chemically perturbed. 
At time $t = 0$, both rotors are collinear with their $\mathbf{t}$ axes pointing along the $z$-direction in the laboratory frame. The rotors are separated by an initial distance $r_0\,a$. The chemical gradient is placed in between the rotors. We explored the response of the rotors to the radial chemical gradient by varying the parameters $\chi$ and $\lambda$ for two different positions of the chemical gradient.

\begin{figure*}[htb]
\centering
\includegraphics[width=\linewidth]{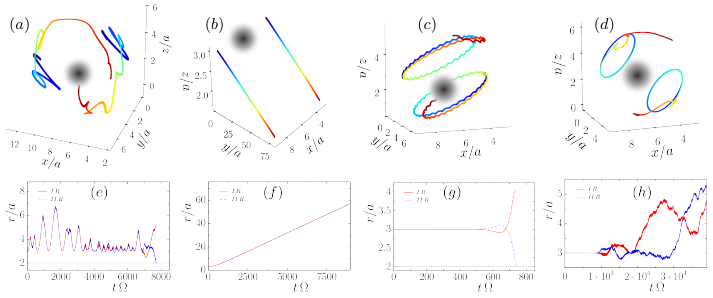}
\caption{
The Top Panel presents the trajectories of the rotors in the presence of a chemical source, and the bottom panel presents the distance of the rotors from the chemical target for the respective trajectories at the top. (a) and (e) is for straight line trajectories (CW-CW) and $\lambda = {v (30,30)}$, $\chi = 6\pi/24$ of first (I R) and second rotor (II R). (b) and (f) is for straight line trajectories (CW-CW) and $\lambda = {v (10,10)}$, $\chi = 9\pi/24$.  (c) and (g) are for circular trajectories (CW-ACW) and $\lambda = {v (-70,70)}$, $\chi = 3\pi/24$. (d) and (h) is for circular trajectories (CW-ACW) and $\lambda = {v (-10,10)}$, $\chi = 7\pi/24$.
The dashed lines in (e) and (g) represent the closest distance to the chemical target reached by the rotor.}
\label{fig:traj r space chem st}
\end{figure*}

\begin{figure}[htb!]
\centering
\includegraphics[scale=1.25]{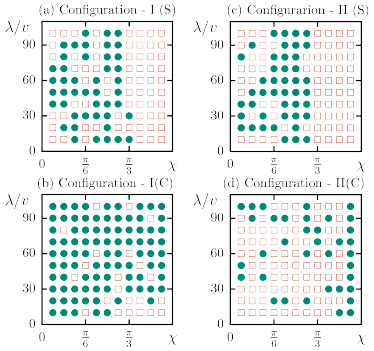}
\caption{(a)-(b) $\lambda$ {\it vs.} $\chi$
Phase diagram considering Configuration-I for rotors with straight-line (S) motion (CW-CW) and circular (C) motion (CW-ACW). (c)-(d) The same but the configuration-II. 
The solid circles indicate the capture of the rotor by the target (chemotaxis), and the hollow square represents rotors moving away from the target (anti-chemotaxis).}
\label{fig:phase space chem st 102 106}
%\end{FPfigure}
\end{figure}

As shown in section \ref{sec:HI}, in the absence of a chemical field, rotors either move in a straight line or a circular path under the influence of each other's hydrodynamic interactions. However, a chemical gradient around them adds to their motion. Therefore, the chemical perturbations and hydrodynamic interaction lead to some interesting swimming states for the rotors, as depicted in Fig.~\ref{fig:traj r space chem st}. A single rotor will not exhibit any translational motion in a chemical gradient. However, the rotational flow field around the rotor will be modified. 
The presence of a second rotor introduces translational motion, as observed before, and hence, the rotors will respond to the gradient either by moving towards it (chemotaxis) or away from it (anti-chemotaxis). 
Fig.~\ref{fig:traj r space chem st} displays the swimming states of the rotors in a radial chemical field for configuration-I shown in Fig.~\ref{fig:chem rotors}. At first, we place the source between the rotors and then evolve them for a long time ($10^6$ simulation time). No systematic dependence of chemotaxis and anti-chemotaxis on the parameters $\chi$ and $\lambda$ is observed. It is clear from Fig.~\ref{fig:traj r space chem st} that both the circular and the straight line trajectories can show chemotaxis and anti-chemotaxis. Moreover, the location of the chemical source plays a vital role in the response of the rotors to it. 
The absolute distance of the chemical target from the rotors and the position of the source relative to them decides fate. As shown in Fig.~\ref{fig:traj r space chem st} rotors move in a straight line for $\chi=6 \pi/4$ end up reaching the chemical source [(a) and (e)], while for $\chi=9\pi/3$ moves away from the source [(b) and (f)]. 
Similar chemotatic and anti-chemotactic behaviour is also seen in circular trajectories [(c)-(g)\& (d)-(h)]. Fig.~\ref{fig:phase space chem st 102 106} summarises the dependence of $\chi$ and $\lambda$ on the chemotactic behaviour of rotors for two different positions of the chemical source.

\subsection{State diagram for chemo and anti-chemotaxis}

It is interesting to see that the behaviour of rotors is not deterministic. 
Most swimming states above $\chi = \pi /3$ for the straight-line trajectories are anti-chemotactic, and the majority below $\chi = \pi/3$ are chemotactic. Note that both configurations show the same behavior. The flow field generated by the rotors for $\chi < \pi/3$ favors the capture. However, the phase diagram for circular trajectories is completely different for the two configurations. The co-linear configuration favours more capture than the non-co-linear state. Therefore, the fate of the rotors depends on multiple factors like their orientation and distance with respect to the source and their trajectories.

\section{Conclusion}
\label{sec:Concluion}

With this work, we have investigated the hydrodynamic behavior of two chiral rotors and their combined response to an external chemical gradient.
We used the generalized squirmer model, called the chiral squirmer, to study the chiral rotors. 
The chiral rotor contributes only from the azimuthal component of the slip velocity. 
In other words, the chiral rotor can generate rotational flows only. 
A single active rotor is stationary in an unbounded fluid, but a pair of rotors may move in either straight line paths or circular paths depending on the nature of the rotors, i.e., whether the rotors are the same kind (same chirality) or opposite (different chirality).  
In the combined behavior of two rotors, the angle $\chi$ and the velocity field strength determined by $\lambda$ play a vital role. By varying these variables, rotors exhibit either linear or circular movements.   The nature of the trajectory depends on the chirality of the rotors, but the direction of the collective motion of the rotors depends on the relative orientation of the rotation axes $(\chi)$. The rotors are always in motion except $\chi = 0, \pi/2$.
Further, we investigated the response of the rotors to external chemical gradients.
We have considered the radial chemical field in the current study as a test case.

The response of a pair of chiral rotors in a chemical landscape is fascinating since they do not possess any intrinsic directional movement but propel with the help of the neighbor's flowfield.  
Depending on various swimming parameters, the former flowfield may drive them toward the chemical target (chemotaxis) or away from it (anti-chemotaxis).
Notably, the initial configurations also control the fate of rotors in a chemical field. While the closed trajectories are prone to show successful chemotaxis in the case of configuration-I, the open trajectories are more successful in showing chemotaxis in configuration-II. 
This work may find its application in designing artificial swimmers to perform manipulated tasks, such as controlled collective motion in a complex environment.

\section{Acknowledgement}
We acknowledge the HPC facility at IISER Bhopal for allocating the necessary computational resources. We additionally acknowledge support from the Austrian Science Fund under project No. ESP 382-N (ESPRIT). The computational results were achieved using the Vienna Scientific Cluster.

\end{document}